\documentclass{article}
\usepackage{lmodern}
\usepackage{tikz}
\usetikzlibrary{arrows,arrows.meta}
\usepackage{amsmath}
\usepackage{scalerel}
\usepackage{mathtools,amssymb,stmaryrd,bm}
\usepackage[T1]{fontenc}
\usepackage{lmodern}
\usepackage{footnote}
\usepackage{fullpage}
\usepackage{hyperref}
\usepackage{bookmark}
\usepackage{color}
\usepackage{graphicx, float}
\usepackage{chemfig}
\usepackage{chemformula}
\usepackage[all,cmtip]{xy}
\usepackage{prftree}

\usepackage{amsthm, url, enumitem}
\usepackage{graphicx}
\usepackage[preview,tightpage]{preview}
\usepackage[]{algorithm2e}
\usepackage{subcaption}
\usepackage[font={md}, labelfont=bf, textfont=it]{caption}

\usepackage{bm}

\newcommand{\abs}[1]{\left| #1 \right|}
\newcommand{\xynode}[1]{*++[F-:<6pt>]{#1}}
\newcommand{\xymedge}[1]{*++[F-]{#1}}

\newcommand{\conc}[1]{\left[ #1 \right]}
\newcommand{\conch}[1]{\left[\ch{#1}\right]}
\newcommand{\conct}[1]{\left[ \mathrm{#1} \right]}
\newcommand{\unit}[0]{\mathbf 0}
\newcommand{\multiset}[1]{\scaleleftright[1.5ex]{\Lbag}{#1}{\Rbag}}
\newcommand{\onenorm}[1]{\left\lVert #1 \right\rVert_1}

\newcommand{\para}[0]{\mathbin{|}}
\newcommand{\apar}[0]{\mathbin{\parallel}}
\newcommand{\ppar}[0]{\parallel}

\mathchardef\mhyphen="2D 
\newcommand{\atrate}[0]{\mathrel{@}}
\newcommand{\trans}[4]{\xymatrix{#1 \ar@{->}[r]^-{#2}_-(1.1){#3} &\;\; #4}}
\newcommand{\transl}[5]{\xymatrix@C=#5em{#1 \ar@{->}[r]^-{#2}_-(1.1){#3} &\;\; #4}}
\newcommand{\transb}[3]{\!\!\xymatrix@C=1.0em{#1 \ar@{->}[r]^-{#2} & #3 }\!\!}
\newcommand{\transbb}[3]{\!\!\xymatrix@C=1.5em{#1 \ar@{->}[r]^-{#2} & #3 }\!\!}
\newcommand{\transbbb}[3]{\!\!\xymatrix@C=2.0em{#1 \ar@{->}[r]^-{#2} & #3 }\!\!}
\newcommand{\transbl}[4]{\!\!\xymatrix@C=#4em{#1 \ar@{->}[r]^-{#2} & #3 }\!\!}
\newcommand{\T}[2]{\mathbf I\!\left(#1 \to #2\right)}
\newcommand{\I}[1]{\mathbf I\!\left( #1 \right)}
\newcommand{\E}[3]{\mathbf E\!\left( #1 \to #2,\, #3 \right)}

\newcommand{\emb}[1]{\left\langle #1 \right\rangle}

\newcommand{\basisc}[3]{\mathbf I\!\left(\!\!\xymatrix@C=1.0em{#1 \ar@{->}[r]^-{#2} & #3 }\!\!\right)}
\newcommand{\basisl}[4]{\mathbf I\bigl(\!\!\xymatrix@C=#4em{#1 \ar@{->}[r]^-{#2} & #3 }\!\!\bigr)}

\makeatletter
\definearrow9{-X>}{%
  \CF@arrow@shift@nodes{#7}%
  \expandafter\draw\expandafter[\CF@arrow@current@style](\CF@arrow@start@node)--(\CF@arrow@end@node)node[midway](Xarrow@arctangent){};%
  \CF@ifempty{#8}
  {\def\CF@Xarrow@radius{0.333}}
  {\def\CF@Xarrow@radius{#8}}%
  \CF@ifempty{#9}%
  {\def\CF@Xarrow@absangle{60}}
  {\pgfmathsetmacro\CF@Xarrow@absangle{abs(#9)}}
  \edef\CF@tmp@str{[\CF@ifempty{#1}{draw=none}{\unexpanded\expandafter{\CF@arrow@current@style}},-]}%
  \expandafter\draw\CF@tmp@str (Xarrow@arctangent)%
  arc[radius=\CF@compound@sep*\CF@current@arrow@length*\CF@Xarrow@radius,start angle=\CF@arrow@current@angle-90,delta angle=-\CF@Xarrow@absangle]node(Xarrow1@start){};
  \edef\CF@tmp@str{[\CF@ifempty{#2}{draw=none}{\unexpanded\expandafter{\CF@arrow@current@style}},-CF]}%
  \expandafter\draw\CF@tmp@str (Xarrow@arctangent)%
  arc[radius=\CF@compound@sep*\CF@current@arrow@length*\CF@Xarrow@radius,%
  start angle=\CF@arrow@current@angle-90,%
  delta angle=\CF@Xarrow@absangle]%
  node(Xarrow1@end){};
  \edef\CF@tmp@str{[\CF@ifempty{#4}{draw=none}{\unexpanded\expandafter{\CF@arrow@current@style}},-]}%
  \expandafter\draw\CF@tmp@str (Xarrow@arctangent)%
  arc[radius=\CF@compound@sep*\CF@current@arrow@length*\CF@Xarrow@radius,start angle=\CF@arrow@current@angle+90,delta angle=\CF@Xarrow@absangle]node(Xarrow2@start){};
  \edef\CF@tmp@str{[\CF@ifempty{#5}{draw=none}{\unexpanded\expandafter{\CF@arrow@current@style}},-CF]}%
  \expandafter\draw\CF@tmp@str (Xarrow@arctangent)%
  arc[radius=\CF@compound@sep*\CF@current@arrow@length*\CF@Xarrow@radius,%
  start angle=\CF@arrow@current@angle+90,%
  delta angle=-\CF@Xarrow@absangle]%
  node(Xarrow2@end){};
  \pgfmathsetmacro\CF@tmp@stra{\CF@Xarrow@radius*cos(\CF@arrow@current@angle)<0?"-":"+"}%
  \pgfmathsetmacro\CF@tmp@strb{\CF@Xarrow@radius*cos(\CF@arrow@current@angle)<0?"+":"-"}%
  \ifdim\CF@Xarrow@radius pt>\z@
    \CF@arrow@display@label{#1}{0}\CF@tmp@stra{Xarrow1@start}{#2}{1}\CF@tmp@stra{Xarrow1@end}%
    \CF@arrow@display@label{#4}{0}\CF@tmp@strb{Xarrow2@start}{#5}{1}\CF@tmp@strb{Xarrow2@end}%
    \CF@arrow@display@label{#3}{0.5}\CF@tmp@stra\CF@arrow@start@node{}{}{}\CF@arrow@end@node%
    \CF@arrow@display@label{#6}{0.5}\CF@tmp@strb\CF@arrow@start@node{}{}{}\CF@arrow@end@node%
  \else
    \CF@arrow@display@label{#2}{0}\CF@tmp@stra{Xarrow1@start}{#1}{1}\CF@tmp@stra{Xarrow1@end}%
    \CF@arrow@display@label{#5}{0}\CF@tmp@strb{Xarrow2@start}{#4}{1}\CF@tmp@strb{Xarrow2@end}%
    \CF@arrow@display@label{#3}{0.5}\CF@tmp@stra\CF@arrow@start@node{}{}{}\CF@arrow@end@node%
    \CF@arrow@display@label{#6}{0.5}\CF@tmp@strb\CF@arrow@start@node{}{}{}\CF@arrow@end@node%
  \fi
}
\definearrow9{<=X>}{%
  \path[allow upside down](\CF@arrow@start@node)--(\CF@arrow@end@node)%
    node[pos=0,sloped,yshift=1pt](\CF@arrow@start@node @u0){}%
    node[pos=0,sloped,yshift=-1pt](\CF@arrow@start@node @d0){}%
    node[pos=1,sloped,yshift=1pt](\CF@arrow@start@node @u1){}%
    node[pos=1,sloped,yshift=-1pt](\CF@arrow@start@node @d1){};%
  \begingroup
    \pgfarrowharpoontrue
    \expandafter\draw\expandafter[\CF@arrow@current@style](\CF@arrow@start@node @u0)--(\CF@arrow@start@node @u1)node[midway](XarrowT@arctangent){};%
    \expandafter\draw\expandafter[\CF@arrow@current@style](\CF@arrow@start@node @d1)--(\CF@arrow@start@node @d0)node[midway](XarrowB@arctangent){};%
  \endgroup
  \CF@arrow@shift@nodes{#7}%
  \CF@ifempty{#8}
  {\def\CF@Xarrow@radius{0.333}}
  {\def\CF@Xarrow@radius{#8}}%
  \CF@ifempty{#9}%
  {\def\CF@Xarrow@absangle{60}}
  {\pgfmathsetmacro\CF@Xarrow@absangle{abs(#9)}}
  \edef\CF@tmp@str{[\CF@ifempty{#1}{draw=none}{\unexpanded\expandafter{\CF@arrow@current@style}},-]}%
  \expandafter\draw\CF@tmp@str (XarrowT@arctangent)%
  arc[radius=\CF@compound@sep*\CF@current@arrow@length*\CF@Xarrow@radius,start angle=\CF@arrow@current@angle-90,delta angle=-\CF@Xarrow@absangle]node(Xarrow1@start){};
  \edef\CF@tmp@str{[\CF@ifempty{#2}{draw=none}{\unexpanded\expandafter{\CF@arrow@current@style}},-CF]}%
  \expandafter\draw\CF@tmp@str (XarrowT@arctangent)%
  arc[radius=\CF@compound@sep*\CF@current@arrow@length*\CF@Xarrow@radius,%
  start angle=\CF@arrow@current@angle-90,%
  delta angle=\CF@Xarrow@absangle]%
  node(Xarrow1@end){};
  \edef\CF@tmp@str{[\CF@ifempty{#4}{draw=none}{\unexpanded\expandafter{\CF@arrow@current@style}},-]}%
  \expandafter\draw\CF@tmp@str (XarrowB@arctangent)%
  arc[radius=\CF@compound@sep*\CF@current@arrow@length*\CF@Xarrow@radius,start angle=\CF@arrow@current@angle+90,delta angle=-\CF@Xarrow@absangle]node(Xarrow2@start){};
  \edef\CF@tmp@str{[\CF@ifempty{#5}{draw=none}{\unexpanded\expandafter{\CF@arrow@current@style}},-CF]}%
  \expandafter\draw\CF@tmp@str (XarrowB@arctangent)%
  arc[radius=\CF@compound@sep*\CF@current@arrow@length*\CF@Xarrow@radius,%
  start angle=\CF@arrow@current@angle+90,%
  delta angle=\CF@Xarrow@absangle]%
  node(Xarrow2@end){};
  \pgfmathsetmacro\CF@tmp@stra{\CF@Xarrow@radius*cos(\CF@arrow@current@angle)<0?"-":"+"}%
  \pgfmathsetmacro\CF@tmp@strb{\CF@Xarrow@radius*cos(\CF@arrow@current@angle)<0?"+":"-"}%
  \ifdim\CF@Xarrow@radius pt>\z@
    \CF@arrow@display@label{#1}{0}\CF@tmp@stra{Xarrow1@start}{#2}{1}\CF@tmp@stra{Xarrow1@end}%
    \CF@arrow@display@label{#4}{0}\CF@tmp@strb{Xarrow2@start}{#5}{1}\CF@tmp@strb{Xarrow2@end}%
    \CF@arrow@display@label{#3}{0.5}\CF@tmp@stra\CF@arrow@start@node{}{}{}\CF@arrow@end@node%
    \CF@arrow@display@label{#6}{0.5}\CF@tmp@strb\CF@arrow@start@node{}{}{}\CF@arrow@end@node%
  \else
    \CF@arrow@display@label{#2}{0}\CF@tmp@stra{Xarrow1@start}{#1}{1}\CF@tmp@stra{Xarrow1@end}%
    \CF@arrow@display@label{#5}{0}\CF@tmp@strb{Xarrow2@start}{#4}{1}\CF@tmp@strb{Xarrow2@end}%
    \CF@arrow@display@label{#3}{0.5}\CF@tmp@stra\CF@arrow@start@node{}{}{}\CF@arrow@end@node%
    \CF@arrow@display@label{#6}{0.5}\CF@tmp@strb\CF@arrow@start@node{}{}{}\CF@arrow@end@node%
  \fi
}
\pgfarrowsdeclare{revhalf}{revhalf}%
{\CF@arrow@size\dimexpr2.5pt+2.5\pgflinewidth\relax
\pgfarrowsleftextend{-\CF@arrow@size}\pgfarrowsrightextend{.5\pgflinewidth}}%
{\CF@arrow@size\dimexpr2.5pt+2.5\pgflinewidth\relax
\pgfsetdash{}\z@\pgfsetroundjoin\pgfsetroundcap
\pgfpathmoveto{\pgfpoint\z@\z@}%
\pgfpathlineto{\pgfpoint{-\CF@arrow@size}{-.5\CF@arrow@size}}%
\pgfpathlineto{\pgfpoint{-.5\CF@arrow@size}\z@}%
\pgfpathlineto{\pgfpoint\z@\z@}%
\pgfusepathqfillstroke}
\definearrow9{<X=>}{%
  \path[allow upside down](\CF@arrow@start@node)--(\CF@arrow@end@node)%
    node[pos=0,sloped,yshift=1pt](\CF@arrow@start@node @u0){}%
    node[pos=0,sloped,yshift=-1pt](\CF@arrow@start@node @d0){}%
    node[pos=1,sloped,yshift=1pt](\CF@arrow@start@node @u1){}%
    node[pos=1,sloped,yshift=-1pt](\CF@arrow@start@node @d1){};%
  \begingroup
    \expandafter\draw\expandafter[\CF@arrow@current@style,-revhalf](\CF@arrow@start@node @u1)--(\CF@arrow@start@node @u0)node[midway](XarrowT@arctangent){};%
    \expandafter\draw\expandafter[\CF@arrow@current@style,-revhalf](\CF@arrow@start@node @d0)--(\CF@arrow@start@node @d1)node[midway](XarrowB@arctangent){};%
  \endgroup
  \CF@arrow@shift@nodes{#7}%
  \CF@ifempty{#8}
  {\def\CF@Xarrow@radius{0.333}}
  {\def\CF@Xarrow@radius{#8}}%
  \CF@ifempty{#9}%
  {\def\CF@Xarrow@absangle{60}}
  {\pgfmathsetmacro\CF@Xarrow@absangle{abs(#9)}}
  \edef\CF@tmp@str{[\CF@ifempty{#1}{draw=none}{\unexpanded\expandafter{\CF@arrow@current@style}},-]}%
  \expandafter\draw\CF@tmp@str (XarrowT@arctangent)%
  arc[radius=\CF@compound@sep*\CF@current@arrow@length*\CF@Xarrow@radius,start angle=\CF@arrow@current@angle-90,delta angle=\CF@Xarrow@absangle]node(Xarrow1@start){};
  \edef\CF@tmp@str{[\CF@ifempty{#2}{draw=none}{\unexpanded\expandafter{\CF@arrow@current@style}},-CF]}%
  \expandafter\draw\CF@tmp@str (XarrowT@arctangent)%
  arc[radius=\CF@compound@sep*\CF@current@arrow@length*\CF@Xarrow@radius,%
  start angle=\CF@arrow@current@angle-90,%
  delta angle=-\CF@Xarrow@absangle]%
  node(Xarrow1@end){};
  \edef\CF@tmp@str{[\CF@ifempty{#4}{draw=none}{\unexpanded\expandafter{\CF@arrow@current@style}},-]}%
  \expandafter\draw\CF@tmp@str (XarrowB@arctangent)%
  arc[radius=\CF@compound@sep*\CF@current@arrow@length*\CF@Xarrow@radius,start angle=\CF@arrow@current@angle+90,delta angle=\CF@Xarrow@absangle]node(Xarrow2@start){};
  \edef\CF@tmp@str{[\CF@ifempty{#5}{draw=none}{\unexpanded\expandafter{\CF@arrow@current@style}},-CF]}%
  \expandafter\draw\CF@tmp@str (XarrowB@arctangent)%
  arc[radius=\CF@compound@sep*\CF@current@arrow@length*\CF@Xarrow@radius,%
  start angle=\CF@arrow@current@angle+90,%
  delta angle=-\CF@Xarrow@absangle]%
  node(Xarrow2@end){};
  \pgfmathsetmacro\CF@tmp@stra{\CF@Xarrow@radius*cos(\CF@arrow@current@angle)<0?"-":"+"}%
  \pgfmathsetmacro\CF@tmp@strb{\CF@Xarrow@radius*cos(\CF@arrow@current@angle)<0?"+":"-"}%
  \ifdim\CF@Xarrow@radius pt>\z@
    \CF@arrow@display@label{#1}{0}\CF@tmp@stra{Xarrow1@start}{#2}{1}\CF@tmp@stra{Xarrow1@end}%
    \CF@arrow@display@label{#4}{0}\CF@tmp@strb{Xarrow2@start}{#5}{1}\CF@tmp@strb{Xarrow2@end}%
    \CF@arrow@display@label{#3}{0.5}\CF@tmp@stra\CF@arrow@start@node{}{}{}\CF@arrow@end@node%
    \CF@arrow@display@label{#6}{0.5}\CF@tmp@strb\CF@arrow@start@node{}{}{}\CF@arrow@end@node%
  \else
    \CF@arrow@display@label{#2}{0}\CF@tmp@stra{Xarrow1@start}{#1}{1}\CF@tmp@stra{Xarrow1@end}%
    \CF@arrow@display@label{#5}{0}\CF@tmp@strb{Xarrow2@start}{#4}{1}\CF@tmp@strb{Xarrow2@end}%
    \CF@arrow@display@label{#3}{0.5}\CF@tmp@stra\CF@arrow@start@node{}{}{}\CF@arrow@end@node%
    \CF@arrow@display@label{#6}{0.5}\CF@tmp@strb\CF@arrow@start@node{}{}{}\CF@arrow@end@node%
  \fi
}
\definearrow3{-|}{%
  \begingroup
  \CF@arrow@shift@nodes{#3}%
  \expandafter\draw\expandafter[\CF@arrow@current@style,-|](\CF@arrow@start@node)--(\CF@arrow@end@node);%
  \CF@arrow@display@label{#1}{0.5}+\CF@arrow@start@node{#2}{0.5}-\CF@arrow@end@node
  \endgroup
}
\makeatother

\newtheorem{thm}{Theorem}[section]

\newtheorem{prop}[thm]{Proposition}

\theoremstyle{definition}

\theoremstyle{definition}
\newtheorem{dfn}[thm]{Definition}
\theoremstyle{remark}

\theoremstyle{remark}

\numberwithin{equation}{subsection}

\title{The Bond-Calculus: A Process Algebra for Complex Biological Interaction Dynamics}

\author{Thomas Wright, Ian Stark \\\\
Laboratory  for  Foundations  of  Computer  Science,\\
School of Informatics, University of Edinburgh, UK\\
\{thomas.wright, ian.stark\}@ed.ac.uk}

\begin{document}
\maketitle

\abstract{
We present the bond-calculus, a process algebra for modelling biological and chemical systems featuring nonlinear dynamics, multiway interactions, and dynamic bonding of agents. Mathematical models based on differential equations have been instrumental in modelling and understanding the dynamics of biological systems. Quantitative process algebras aim to build higher level descriptions of biological systems, capturing the agents and interactions underlying their behaviour, and can be compiled down to a range of lower level mathematical models. The bond-calculus builds upon the work of Kwiatkowski, Banks, and Stark's continuous pi-calculus by adding a flexible multiway communication operation based on affinity patterns and general kinetic laws. We develop a compositional semantics based on vector fields and linear operators, which we use to define the time evolution of this system. This enables simulation and analysis via differential equation generation or stochastic simulation. Finally, we apply our framework to an existing biological model: Kuznetsov's classic model of tumour immune interactions.
}

\maketitle

\tableofcontents

\section{Introduction}
\label{introduction}
A variety of quantitative process algebras such as the stochastic $\pi$-calculus~\cite{Priamistochpi}, the continuous $\pi$-calculus~\cite{kwiatkowski/stark:continuous-pi}, and Bio-PEPA~\cite{biopepapaper} have developed over the last decade to provide high level modelling languages, capable of describing a wide range of complex biochemical systems including signalling networks and gene regulatory networks as interactions between a number of \emph{agents} or \emph{processes}. By viewing systems as collections of distinct agents, whose behaviour emerges from the interactions between them, process algebra adopts a \emph{interaction centric} viewpoint, whilst enforcing the principle of \emph{compositionality}: the behaviour of a system is built from the behaviour of its components~\cite{priami2004biosystems}. Process algebra models are amenable to simulation and analysis using both discrete, stochastic methods based on Gillespie's Stochastic Simulation Algorithm~\cite{gillespie1977exact}, and continuous, deterministic methods via a fluid/mean field approximation of the dynamics as a system of differential equations, decoupling the method of analysis from the model representation. However, most existing biological process algebras are split between languages such as Bio-PEPA which uses a variant of CSP style synchronisation~\cite{Hoare2002} to capture complex features of chemical reactions such as multi-way interactions, stoichiometry, and general kinetic laws~\cite{biopepapaper} but cannot describe dynamically created processes, and languages such as stochastic $\pi$~\cite{Priamistochpi} and continuous $\pi$~\cite{kwiatkowski/stark:continuous-pi} which use \emph{name passing} or \emph{mobility} (where processes exchange channel names over channels/sites) to model bonding of agents to dynamically form new agents, but are limited to modelling binary, mass action reactions~\cite{calder2009modellingstyles}. Mobility makes it possible to model some of the most interesting features of biochemical systems such as dynamic network topologies, membrane interactions~\cite{aman2011mobility}, and chemical bonding~\cite{regev2001} motivating us to develop a language which better combines mobility with the other key features of biochemical networks and to develop a mathematical semantics capable of backing up such an expressive quantitative language.

The main contribution of this paper is a new process algebra, the \emph{bond-calculus}, which aims to bridge the divide by effectively modelling the full range of biochemical reactions featuring multi-way reactions, dynamic bonding of agents, and general kinetics. This is made possible via a distinctive communication operation, \emph{open multi-way bonding}, in which parties may join a multi-way reaction one by one whilst agreeing on a number of channel/site names to establish bonds. The resulting reactions will be atomic events, with overall rate determined compositionally based on the concentrations of the participants and a general kinetic law. The language also introduces pattern matching against a network of \emph{affinity patterns}, which generalises the notion of reaction site affinity to provide a flexible way of specifying complex multi-molecular reactions and their rates. Similarly to Rule Based Modelling~\cite{danos2004formal}, affinity patterns make it possible to define general patterns of interaction (and their kinetics) which are reused by many different reactants, however, the bond-calculus is an agent based language more similar to traditional process calculi. The language is equipped with a compositional semantics in terms of vector fields and linear operators, which provides a means of generating a system of Ordinary Differential Equations (ODEs) governing the dynamics of a given model. We demonstrate the language, through a model of tumour immune cell interactions featuring dynamic bonding and multiway interactions, based on at Kuznetsov's classic immunogenic tumour growth model~\cite{kuznetsov1994tumourmodel}.

Throughout the rest of this section we will introduce the main features of the languages through some examples of enzymatic reactions. In Section~\ref{sec:relatedwork} we will discuss related work and how the bond-calculus relates to existing formal biochemical modelling languages. In Section~\ref{sec:semantics} we briefly describe the semantics of the language and how we simulate its dynamics. Finally, in Section~\ref{sec:modellingexamples} we apply the bond-calculus to capture Kuznetsov's model of immunogenic tumour growth~\cite{kuznetsov1994tumourmodel}.

\begin{figure}
  \centering
  \begin{subfigure}{0.4\textwidth}
    \centering
    \schemestart
    $S$ \arrow{<=X>[$E$][][$k_1$][][][$k_{-1}$]} $C$
        \arrow{-X>[][$E$][$k_2$]}$P$
        \arrow{->[$k_3$]}
    \schemestop
    \caption{Mass action kinetics}
    \label{enzymema}
  \end{subfigure}%
  \begin{subfigure}{0.4\textwidth}
    \vspace{0.1em}
    \centering
    \schemestart
    $S$ \arrow{-U>[$E$][$E$]} $P$
        \arrow{->[$k_3$]}
    \schemestop
    \vspace{0.8em}
    \caption{Michaelis-Menten kinetics}
    \label{enzymemm}
  \end{subfigure}
  \begin{subfigure}{0.7\textwidth}
    \centering
    \vspace{0.7em}
    \includegraphics[width=\textwidth]{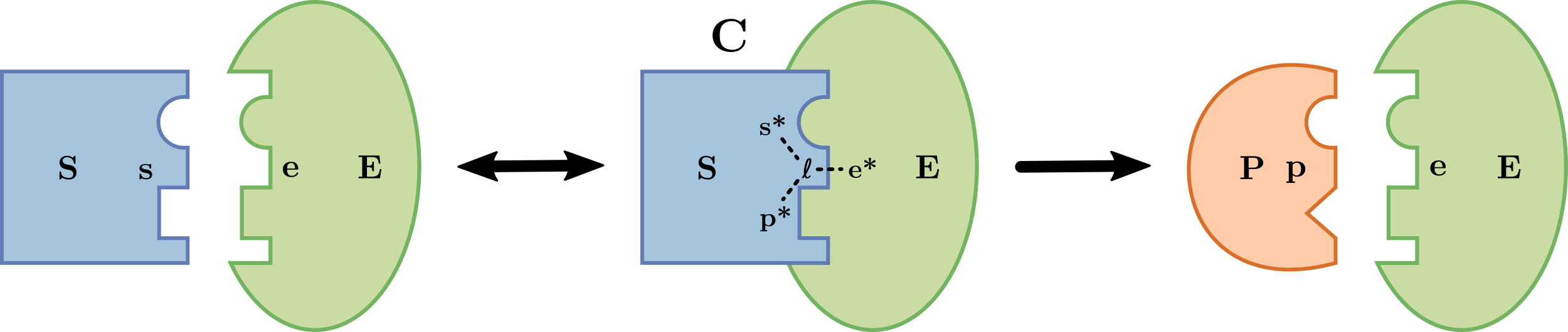}
    \caption{The sites and locations in the enzyme binding process.}
    \label{enzymesites}
  \end{subfigure}
  \caption{Simple enzymatic reaction}
\end{figure}

\paragraph{Species, sites and locations}
A bond-calculus model consists of a \emph{mixture} $\conc{A_1} A_1 \ppar \ldots \ppar \conc{A_n} A_n$ of different \emph{species} $A_1,\ldots,A_n$ at given continuous concentrations $\conc{A_1},\ldots,\conc{A_n} \in \mathbb R_{\geq 0}$. Each species represents the behaviour of a single molecule of a chemical species and the overall behaviour of the mixture is determined by the collective behaviour and interactions of its species. A species has several \emph{sites} ($s, e, p, \ldots $) at which it may engage in reactions with other molecules at compatible sites (see, for example, Figure~\ref{enzymesites}). Sites may also be annotated with locations $\ell,m,\ldots$, by writing $s@\ell, e@m, \ldots$, recording the site's internal location within a molecule -- sites at the same location of a molecule may interact allosterically. The combinations of sites which are compatible and may engage in a reaction are recorded by an \emph{affinity pattern}. For example, the pattern $s \apar e$ indicates that a molecule with site $s$ will react with a molecule with site $e$, and will match a mixture of form $\ldots \ppar \conc{S} S \ppar \conc{E} E \ppar \ldots$ where species $S$ has site $s$ and species $E$ has site $e$ (the order of species in the mixture does not matter). Overall, a bond-calculus model must specify the following four components: 
\begin{itemize}
  \item A set of \emph{species definitions} $A \triangleq \ldots$ which define the behaviour of each chemical species as an agent of the process calculus. The species are defined by composing existing species using parallel composition $|$, choice $+$, by location definitions $(\nu\,\ell_1,\ldots,\ell_n)S_{\ell_1,\ldots,\ell_n}$ which define new bound locations $\ell_1,\ldots,\ell_n$ in species $S_{\ell_1,\ldots,\ell_n}$, or by communication prefixes of form $s.S$, $s@\ell.S$, $s(\ell).S_\ell$, or in general $s@\ell(m_1,\ldots,m_n).S_{m_1,\ldots,m_n}$ which defines a species which has a reaction site $s$ at location $\ell$, and upon reaction on site $s$ receives locations $m_1,\ldots,m_n$ to become species $S_{m_1,\ldots,m_n}$.

  \item A \emph{mixture definition} $\Pi \triangleq \conc{A_1} A_1 \ppar \ldots \ppar \conc{A_n} A_n$ specifying that the model consists of chemical species $A_1,\ldots,A_n$ at starting concentrations $\conc{A_1},\ldots,\conc{A_n} \in \mathbb R_{\geq 0}$ .

  \item A set of kinetic law definitions $L(x_1,\ldots,x_m) \triangleq \ldots$ specifying the different rate laws which may be used in reactions. When the rate law is applied to a reaction, the arguments $x_1,\ldots,x_m$ will be the concentrations of each reacting site.

  \item An affinity network $\mathcal A \triangleq \{\,\bm\gamma_1 \atrate L_1, \ldots,\bm\gamma_n \atrate L_n\,
  \}$ with affinity patterns $\bm\gamma_j$ specifying the available reactions, and kinetic laws $L_j$ specifying the reaction rates.
\end{itemize}

\begin{figure}[b]
  \[
    \xymatrix{
      \xynode{s} \ar@{-}[r] &
      \xymedge{\operatorname{MM}_{V_{\mathrm{max}},k}} \ar@{-}[r] &
      \xynode{e} &
      \xynode{p} \ar@{-}[r] &
      \xymedge{\mathrm{MA}_k}
    }
  \]%
  \caption{The affinity network $\mathcal A_{\operatorname{MM}}$ as a hypergraph.}
  \label{fig:affinity-network-hypergraph}
\end{figure}
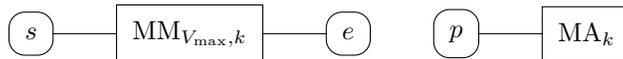

For example, consider the enzymatic reaction in Figure~\ref{enzymema}, in which substrate $S$ is transformed into product $P$ via the intermediate complex $C$ formed by binding with enzyme $E$ ($P$ is also assumed to decay at rate $k_3$). This can be approximately modelled as a single reaction using the Michaelis-Menten kinetic law~\cite{michaelis1913kinetik} as in Figure~\ref{enzymemm}. This model can be described in the bond-calculus as the mixture
\[\Pi \triangleq \conc{S} S \ppar \conc{E} E \ppar \conc{P} P\]
with the species defined by
\begin{align*}
  S & \triangleq s.P & E & \triangleq e.E & P & \triangleq p.\unit 
\end{align*}
Here we have a mixture of species $S$ and $E$ where $S$ is defined as the prefix $s.P$ specifying that it has a single reaction site $s$ and can turn into $P$, and similarly $E$ has a single site $e$ and can turn into $E$  (that is, $E$ is unchanged by reactions on $e$), whilst $P$ has a single site $p$ and decays into the null species $\unit$. The reaction between $S$ and $E$ is governed by the Michaelis-Menten kinetic law,
\begin{align*}
  \operatorname{MM}_{V_{\mathrm{max}},k}(\conc{S}, \conc{E})
  \triangleq
  \frac{V_{\mathrm{max}} \conc{S}\!\conc{E}}{k + \conc{E}}
\end{align*}
whilst the decay of $P$ is governed by the mass action kinetic law (in the unary case of $n=1$),
\begin{align*}
  \operatorname{MA}_k(\conc{X_1},\ldots,\conc{X_n})
  \triangleq
  k \conc{X_1} \ldots \conc{X_n}
  .
\end{align*}
We can specify these reactions via the affinity network,
\begin{align*}
  \mathcal A_{\operatorname{MM}} \triangleq \bigl\{\, s \apar e \atrate \operatorname{MM}_{V_{\mathrm{max}},k},\,
                                  p \atrate \operatorname{MA}_{k_3} \,\bigr\}
\end{align*}
which consists of affinity patterns $s \apar e \atrate \operatorname{MM}_{V_{\mathrm{max}},k}$ specifying that site $s$ reacts with site $e$ at rate $\operatorname{MM}_{V_{\mathrm{max}},k}(\conc{s}, \conc{e}) = \frac{V_{\mathrm{max}} \conc{s}\!\conc{e}}{k + \conc{e}}$ (where $\conc{s}$ and $\conc{e}$ are the total concentrations of sites $s$ and $e$ respectively), and $p \atrate \operatorname{MA}_{k_3}$ specifying that site $p$ engages in a unary reaction at rate $\operatorname{MA}_{k_3}(\conc{p}) = k_3 \conc{p}$.
The affinity network may also be viewed graphically as a hypergraph with nodes labelled with sites and hyperedges labelled with kinetic laws as shown in Figure~\ref{fig:affinity-network-hypergraph} -- in this way our affinity networks extend the affinity networks of continuous~$\pi$~\cite{kwiatkowski/stark:continuous-pi} which are restricted to graphs with edges labelled by stoichiometric rates.

\paragraph{Dynamic bonding}
It is also possible to model the enzyme reaction directly using mass action kinetics where $S$ and $E$ bond dynamically to form the complex $C$ as in Figure~\ref{enzymemm}. In dynamic bonding, two or more agents will react at compatible sites ($s,e$) and agree upon a number of new shared locations ($\ell$) -- they will then form a complex of several agents joined in parallel composition $|$ which are bound together by their shared locations (as shown in Figure~\ref{enzymesites}). This can be done with species,
\begin{align*}
  S & \triangleq s(\ell).(s^*@\ell.S + p^*@\ell.P) & 
  E & \triangleq e(\ell).e^*@\ell.E &
  P & \triangleq p.\unit
\end{align*}
where $S$ communicates at site $s$ to receive location $\ell$ to become $S^*_\ell \triangleq s^*@\ell.S + p^*@\ell.P$, whilst $E$ communicates at $e$ to receive location $\ell$ to become $E^*_\ell \triangleq e^*@\ell.E$. Thus, together they form the complex,
\begin{align*}
  C \triangleq (\nu\,\ell)(S^*_\ell \para E^*_\ell)
    = (\nu\,\ell)\!\left(
      (s^*@\ell.S + p^*@\ell.P)\para e^*@\ell.E
  \right).
\end{align*}
This includes the restriction $(\nu\,\ell)(\ldots)$ which binds an internal location name $\ell$ within the complex -- then sites $s@\ell$, $p^*@\ell$, $e^*@\ell$ are placed at location $\ell$ specifying they all lie at this internal location $\ell$ within the complex and may interact allosterically together. The parallel composition $|$ joins the components of the complex coming from $S$ and $E$ respectively, and they are bound together in a single molecule due the shared internal location $\ell$. The affinity network of this model is given by,
\begin{align*}
  \mathcal A_{\operatorname{MA}} \triangleq
  \bigl\{\,
    s \apar e \atrate \operatorname{MA}_{k_1},\,
    s^*|e^* \atrate \operatorname{MA}_{k_{-1}},\,
    p^*|e^* \atrate \operatorname{MA}_{k_{2}},\,
    p \atrate \operatorname{MA}_{k_3}
  \,\bigr\}.
\end{align*}
The network includes the affinity patterns $s \apar e \atrate \operatorname{MA}_{k_1}$ specifying that sites $s$ and $e$ react at rate $\operatorname{MA}_{k_1} $, $s^*|e^* \atrate \operatorname{MA}_{k_{-1}}$ specifying that sites $s^*$ and $e^*$ can interact internally to a molecule leading to a reaction at rate $\operatorname{MA}_{k_{-1}}$ (causing the complex $C$ to break back down into $S$ and $E$ at rate $k_{-1} \conc{C}$), and $p^*|e^* \atrate \operatorname{MA}_{k_{2}}$ specifying that sites $s^*$ and $e^*$ can interact internally to a molecule leading to a reaction at rate $\operatorname{MA}_{k_{2}}$ (causing the complex $C$ to transform into $P$ and $E$ at rate $k_2 \conc{C}$).

\paragraph{Multi-way reactions}

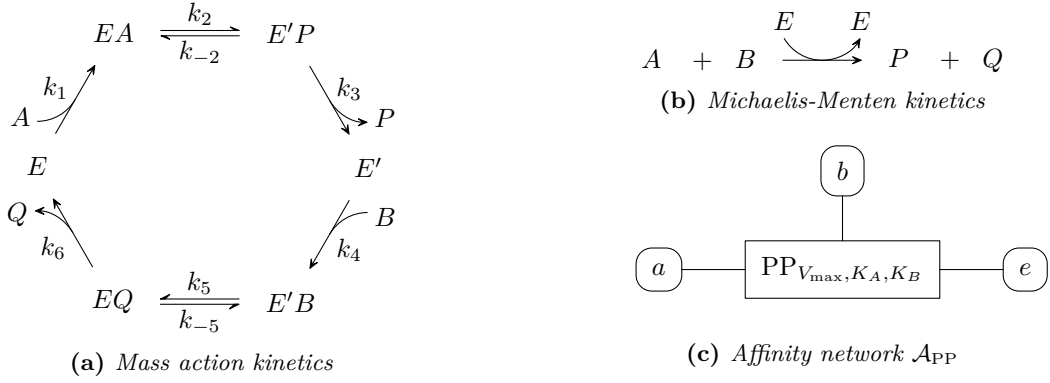
\begin{figure}
  \centering
  \begin{subfigure}{0.4\textwidth}
  \centering
  \schemestart
    $E$ \arrow{-X>[*{0.east}$A$][][*{0}$k_1$]}[60]
    $EA$ \arrow{<=>[$k_2$][$k_{-2}$]}
    $E'P$ \arrow{-X>[][*{0.west}$P$][*{0}$k_3$]}[-60]
    $E'$ \arrow{-X>[*{0.west}$B$][][*{0.north west}$k_4$]}[-120]
    $E'B$ \arrow{<X=>[][][$k_{-5}$][][][$k_5$]}[-180]
    $EQ$ \arrow{-X>[][*{0.east}$Q$][*{0.north east}$k_6$]}[-240]
  \schemestop
  \caption{Mass action kinetics}
  \label{pingpongma}
  \end{subfigure}%
  \begin{subfigure}{0.6\textwidth}
  \centering
  \begin{subfigure}{0.5\textwidth}
  \centering
  \schemestart
    $A$ \+ $B$ \arrow{-U>[$E$][$E$]} $P$ \+ $Q$
  \schemestop
  \caption{Michaelis-Menten kinetics}
  \label{pingpongmm}
  \end{subfigure}

  \begin{subfigure}{0.5\textwidth}
  \[
   \xymatrix@R=17pt{
      & \xynode{b} \ar@{-}[d] \\
      \xynode{a} \ar@{-}[r] &
      \xymedge{\operatorname{PP}_{V_\mathrm{max}, K_A, K_B}} &
      \xynode{e} \ar@{-}[l]
    } 
  \]
  \caption{Affinity network $\mathcal A_{\mathrm{PP}}$}
  \end{subfigure}
  \end{subfigure}
  \caption{The Ping-Pong Mechanism}
\end{figure}
It is common to model more complex enzymatic reactions, as a single multi-way reaction governed by a general kinetic laws. One example is given be the Ping-Pong mechanism (Figure~\ref{pingpongma}) in which reactions $A \to P$ and $B \to Q$ are catalysed via a single enzyme with two states $E$ and $E'$. In the first stage of the reaction $A$ is transformed into $P$ whilst the enzyme $E$ acts as a catalyst and is turned into $E'$, which in turn acts as a catalyst for the second stage in which $B$ is turned into $Q$ and $E'$ is turned back into $E$. In this way a dynamic balance of both states of the enzyme are maintained, allowing both stages of the reaction to progress. This scheme can be modelled by the single ternary reaction in Figure~\ref{pingpongmm} with the Michaelis-Menten kinetic law,
\[
  \operatorname{PP}_{V_\mathrm{max}, K_A, K_B}(\conc{A}, \conc{B}, \conc{E})
  \triangleq
  \frac{V_\mathrm{max} \conc{A} \!\conc{B} \! \conc{E}}
       {K_A \conc{B} + K_B \conc{A} + \conc{A} \!\conc{B}}
  \text{.}
\]
This reaction can then be modelled with the species,
\begin{align*}
  A & \triangleq a.P &
  B & \triangleq b.Q &
  E & \triangleq e.E
  \\
  P & \triangleq p.P &
  Q & \triangleq q.Q,
\end{align*}
and the affinity network,
\[
  \mathcal A_{\operatorname{PP}} \triangleq \bigl\{
    a \apar b \apar e
    \atrate \operatorname{PP}_{V_\mathrm{max}, K_A, K_B} 
  \bigr\}
\]
where the affinity pattern $a \apar b \apar e
    \atrate \operatorname{Ping-Pong}_{V_\mathrm{max}, K_A, K_B}$ specifies a ternary reaction between sites $a$, $b$, and $e$ at rate $\operatorname{Ping-Pong}_{V_\mathrm{max}, K_A, K_B}(\conc{a},\conc{b},\conc{e})$.

\paragraph{Symmetric multi-way bonding}

\begin{figure}
  \begin{subfigure}{0.33\textwidth}
    \centering
    \vspace{32px}
    \includegraphics[height=48px]{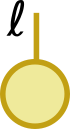}
    \begin{align*}
    A & \triangleq a(\ell).A^*_{\ell} \\
    A^*_{\ell} & \triangleq a^*@\ell.A
    \end{align*}%
    \vspace{-1.5em}
    \caption{Monomer}
    \label{fig:autocatalysis1}
  \end{subfigure}%
  \begin{subfigure}{0.33\textwidth}
    \centering
    \includegraphics[height=80px]{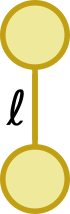}
    \begin{align*}
      \mathcal A_2 \triangleq \bigl\{\,
        &a \apar a \hspace{0.247em} \atrate \operatorname{MA}_{k_2},\\
        &a^* | a^* \atrate \operatorname{MA}_{k_{-2}}
      \,\bigr\}
    \end{align*}%
    \vspace{-1.5em}
    \caption{Dimer}
    \label{fig:autocatalysis2}
  \end{subfigure}%
  \begin{subfigure}{0.33\textwidth}
    \centering
    \includegraphics[height=80px]{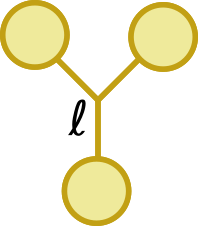}
    \begin{align*}
      \mathcal A_3 \triangleq \bigl\{\,
        & a \apar a \apar a \atrate \operatorname{MA}_{k_3},\\
        & a^* | a^* | a^* \atrate \operatorname{MA}_{k_{-3}}
      \,\bigr\}
    \end{align*}%
    \vspace{-1.5em}
    \caption{Trimer}
    \label{fig:autocatalysis3}
  \end{subfigure}
  \caption{Symmetric bonding at a single site.}
\end{figure}

There is a significant mismatch between traditional binary name-passing process calculi such as the $\pi$-calculus, where communication is directional, from a sender to a receiver, and chemical bonding which is symmetric: multiple parties bond based on the compatibility or affinity of their reaction sites. This problem is most acute in the case of homooligomerization, where multiple identical molecules (monomers) bind to form a single unit. In the $\pi$-calculus this would have to be modelled by agents which non-deterministically choose which role to play (see e.g.~\cite{kuttler2006}), breaking the symmetry of the model. However, the bond-calculus has a completely symmetric communication operation based the $\pi\mathrm{I}$-calculus~\cite{sangiorgi1996pi}, and so can model a monomer with a single autoreactive site with a single, symmetric agent definition (Figure~\ref{fig:autocatalysis1}). This pattern covers, for example, domain swapping~\cite{bennettdomainswapping1995} which explains the symmetric structure of many proteins via a symmetric binding process where monomers exchange identical subunits. Then the affinity network $\mathcal A_2$ (Figure~\ref{fig:autocatalysis2}) has affinity pattern $a \apar a$ which specifies $A$ can form into dimers,
\[
  (\nu\,\ell)(A^*_\ell \para A^*_\ell){}
\]
via an autocatalytic reaction at $a$, and $a^* | a^*$ specifying these dimers may break apart via internal interactions at $a^*$. Moreover, under affinity network $\mathcal A_3$ this extends to $3$-way bonding (Figure~\ref{fig:autocatalysis3}), so three monomers will bond automically into a trimer, 
\[
  (\nu\,\ell)(A^*_\ell \para A^*_\ell \para A^*_\ell).
\]

The reaction rates will be adjusted according to the fact we have only a single site on a single chemical species, so the rate of bonding under $\mathcal A_2$ will be $\frac{1}{2!} \operatorname{MA}_{k_2}(\conc{A},\conc{A}) = \frac{1}{2}k_2\conc{A}^2$, whilst the rate of bonding under $\mathcal A_3$ will be $\frac{1}{3!} \operatorname{MA}_{k_3}(\conc{A},\conc{A},\conc{A}) = \frac{1}{6} k_3 \conc{A}^3$. However, if we had a monomer such as $B \triangleq a.B^*_{\ell} + b.B^{**}_\ell$ which dimerizes via reactions at two different sites which interact according to the pattern $a \ppar b @ \operatorname{MA}_k$, the bonding rate would be $\operatorname{MA}_{k}(\conc{B},\conc{B}) = k\conc{B}^2$. By these combinatorial considerations we are able to calculate the correct rates of reactions no matter whether the sites in a pattern occur in the same species or in different species and regardless of how may times a single site is repeated in an affinity pattern; see the discussions of reaction rates in \cite{cardelli2008processratesemantics,kade2017}.

\paragraph{Overlapping sites and allosteric modification}
\begin{figure}[b]
  \centering
  \begin{subfigure}{0.4\textwidth}
    \centering
    \includegraphics[width=0.8\textwidth]{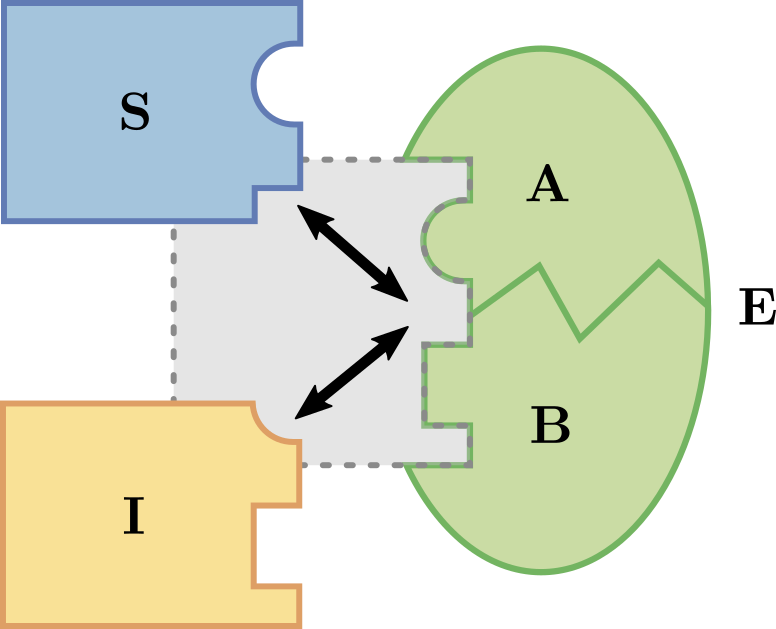}
  \end{subfigure}%
  \begin{subfigure}{0.4\textwidth}
    \centering
    \schemestart
    $E$ \arrow(e--c){<=X>[$S$][][$r_1$][][][$r_{-1}$]} $C$ \arrow{-X>[][$E$][$r_2$]} $P$
    \arrow(@e--ei){<X=>[][][*{0}$r_{-3}$][*{0}$I$][][*{0}$r_3$]}[-90]$D$
    \schemestop
 
  \end{subfigure}
  \caption{Enzyme with inhibition by overlapping sites}
  \label{enzymeoverlappingsites}
\end{figure}
By modelling chemical reaction sites as channels, process algebras assume the different reaction sites on a molecule to be independent. However, in many molecules the reaction capabilities of a site depend on interactions with other sites on a molecule, either in the form of physically overlapping binding sites, or allosteric modification of the conformity of one site via structural changes arising from the binding of another -- these interactions are a key aspect of both protein interactions and gene regulation. The problem of modelling overlapping, mutually exclusive binding sites has been considered by~\cite{kuttler2006,kuttler2007spico}, with existing solutions either relying on atomically combining events via instantaneous actions~\cite{kuttler2006,kuttler2007spico} or transactions~\cite{ciocchetta2007biologicaltransactions}, or relying on updates of shared state via shared attributes in the Attributed $\pi$-calculus~\cite{John2008attributedpi} or via global merging of channels though name fusions~\cite{ciobanu2008,ciobanu2013}. However, in the bond-calculus we can use affinity patterns to model binding and internal interactions as a single multi-way reaction (involving multiple components of a single reactant). For example, we can give a new definition of an enzyme with two mutually exclusive binding sites $A$ and $B$ (as shown in Figure \ref{enzymeoverlappingsites}),
\begin{align*}
  E & \triangleq (\nu\,\ell) (A_\ell \para B_\ell) \\
  A_\ell & \triangleq a@\ell(m).A^*_{\ell,m} + a^*@\ell.A_\ell &
  A^*_{\ell,m} & \triangleq a^{**}@m.A_\ell \\
  B_\ell & \triangleq b@\ell(m).B^*_{\ell,m} + b^*@\ell.B_\ell &
  B^*_{\ell,m} & \triangleq b^{**}@m.B_\ell
\end{align*}
Then if we give the same substrate and product definition as before,
\begin{align*}
  S & \triangleq s(\ell).(s^*@\ell.S + p^*@\ell.P) & 
  P & \triangleq p.\unit
\end{align*}
(where $s$ is assumed to bind to site $a$ of the enzyme), we can introduce an inhibitor,
\begin{align*}
  I & \triangleq i(\ell).I^*_{\ell} &
  I^*_{\ell} & \triangleq i^*@\ell.I
\end{align*}
whose site $i$ binds to site $b$ of the enzyme, stopping it from catalysing the reaction by allosterically modifying the site $a$. We may then define the affinity network,
\begin{align*}
  \mathcal A_{\mathrm{I}} & \triangleq \bigl\{\,
    s \apar a|b^* \atrate \operatorname{MA}_{k_1},\,
    i \apar a^*|b \atrate \operatorname{MA}_{k_3},\,
    s^*|a^{**} \atrate \operatorname{MA}_{k_{-1}},\,
    i^*|b^{**} \atrate \operatorname{MA}_{k_{-3}},\,
    a^{**}|p^* \atrate \operatorname{MA}_{k_{2}}
  \,\bigr\}
\end{align*}
where the affinity pattern $s\apar a|b^* \atrate \operatorname{MA}_{k_1}$ specifies that site $s$ can only react with a molecule which has both site $a$ (causing site $A$ to be become bound) and site $b^*$ (indicating that site $B$ is unbound) whilst site $i$ may only reaction with a molecule with both site $a^*$ (indicating that site $A$ is unbound) and site $b$ (indicating binding at site $B$). Once $S$ and $E$ have bound at site A to form complex,
\[
  C \triangleq (\nu\,\ell,m)(A_{\ell,m} \para B_\ell \para (s^*@m.S + p^*@m.P))
\]
the pattern $s^*|a^{**} \atrate \operatorname{MA}_{k_{-1}}$ specifies they may unbind to release $S$ and $E$ whilst the pattern $a^{**}|p^* \atrate \operatorname{MA}_{k_{2}}$ specifies they may release $P$ and $E$. On the other hand when $E$ and $I$ bind at site $B$ to form the complex 
\[
  D \triangleq (\nu\,\ell,m)(A_{\ell} \para B_{\ell,m} \para I_m)
\]
the pattern $i^*|b^{**} \atrate \operatorname{MA}_{k_{-3}}$ specifies they may unbind to release $S$ and $I$. This example demonstrates the ability of affinity patterns to capture complex multi-molecular with multiple reacting sites both within a single molecule, and spanning different molecules.






\subsection{Related work}
\label{sec:relatedwork}
The design of the bond-calculus was originally conceived as an extension of the continuous $\pi$-calculus to model a broader range of chemical reactions -- continuous~$\pi$~\cite{kwiatkowski/stark:continuous-pi} and BlenX~\cite{Dematte2008BlenXTutorial} both pioneered a form of affinity network, however, both are restricted to unary and binary reactions, continuous $\pi$ to the law of mass action, whilst neither supports the flexible pattern matching of affinity patterns. In~\cite{ciocchetta2007biologicaltransactions} biological transactions were considered as a way of modelling multi-reactant reactions atomically in the stochastic $\pi$-calculus~(and later in~\cite{ciocchetta2008blenXT} for BlenX), however, this required a rather heavyweight language extension and the concept of transactions is somewhat unnatural from a biologically modelling perspective. Bio-PEPA~\cite{biopepapaper} simplified matters by modelling multi-reactant reactions directly as $n$-way synchronisation, however, this approach relied on PEPA/Bio-PEPA's CSP style communication operator, which, unlike $\pi$-calculus style communication, cannot describe dynamic formation of agents/complexes.  Bio-PEPA~\cite{chabrier2003computational} and BlenX~\cite{ciocchetta2008blenXT} both support modelling non-mass action reactions using functional rates, and our approach was particularly influenced by Bio-PEPA and its ODE extraction method~\cite{biopepapaper}. sCCP follows a somewhat different approach, modelling chemical species as a number of variables subject to concurrent constraints, and is able to model arbitrary chemical reactions~\cite{bortolussi2006sccp,Bortolussi2008sccpbio}; attributed process calculi such as Attributed $\pi$~\cite{John2008attributedpi}  and Imperative $\pi$~\cite{sangiorgi1996pi} can model general kinetics in a similar manner~\cite{mazemondet2009imppigeneral}. One significant difference is that whilst in both Bio-PEPA and BlenX functional rates are global expressions which depend directly on the concentrations of certain specific \emph{species}, whereas we model kinetic laws as reusable functions and follow a compositional approach to deriving reaction rates based on the concentrations of each \emph{site} involved in the reaction, which is more similar to the method of deriving reaction rates in PEPA~\cite{tribastone2012scalable,benkirane2009stirlingamendment} (see \cite[Section~1.2]{benkirane2011thesis} for a discussion of the differences in reaction rates in PEPA and Bio-PEPA) -- this better fits our site oriented viewpoint and allows us to model multiple species involved in a reaction at the same site.

One of the main features of the bond-calculus is the use of pattern matching to specify compatible reaction sites in the form of an affinity network. Rule based biochemical modelling languages such as Kappa~\cite{danos2004formal} and BIOCHEM~\cite{fages2002modelling} also specify possible reactions via pattern matching, however, whilst these languages are reaction oriented, the bond-calculus follows an agent based approach with the results of reactions specified by the behaviour of species. This use of pattern matching in an agent based language is similar to the join-calculus'~\cite{fournet2002joincalculus} use of join patterns, however, whilst the join-calculus targets distributed concurrent programming and uses local synchronisation, in the biochemical context it makes more sense to match patterns globally against a shared affinity network. The bio-calculus~\cite{nagasaki1999bio} is a variant of the join-calculus adapted to molecular interactions, however, it uses pattern matching to directly specify chemical reactions rather than site affinity and so its modelling style is closer to chemical reaction networks or rule based models. The bond-calculus' use of sites annotated with location allows a second type of pattern matching and is similar to polyadic synchronisation in $\pi@$~\cite{versari2008piat} and input patterns in \textsc{SpiCO}~\cite{kuttler2007spico}, however, the introduction of a separate class of (internally mobile) locations referring to internal positions on a molecule (as opposed to global locations in a model) is new and in combination with affinity patterns this provides a flexible way to describe internal interactions and molecular binding.

In order to combine name passing with multi-way reactions the bond-calculus uses a symmetric communication operation inspired by Davide Sangiorgi's $\pi\mathrm{I}$-calculus~\cite{sangiorgi1996pi} and its concept of internal mobility, but extended to $n$-way interactions. The use of a symmetric communication operation for a biochemical process algebra has already been explored by continuous~$\pi$~\cite{kwiatkowski/stark:continuous-pi}, by BlenX~\cite{Dematte2008BlenXTutorial}, and by~\cite{duchier2006multibehaviouralconcurrentobjects}, however all of these used a form of simultaneous exchange restricted to binary interactions, which is quite different from the communication operation adopted by the bond-calculus. More similar is the stochastic fusion calculus~\cite{ciobanu2008,ciobanu2012} which has form of symmetric communication based on fusion of names. Whilst name fusion is still a binary communication operation, it results in globally merging the communicated names (unlike $\pi\mathrm{I}$  where names are merged locally) which may impact many different processes; this can be used to model many-to-many communication in cooperative gene regulation~\cite{ciobanu2008,ciobanu2013,ciobanutr}. The link-calculus~\cite{bodei2013linkcalculus} is another approach to extending the $\pi$-calculus to support multi-way interactions, and both Bio-PEPA and the bond-calculus fit its concept of open multi-party interactions. It builds up multi-way interactions as chains of links, and its process communication mechanism is a form of $n$-way pattern matching/simultaneous exchange; this is very flexible, can atomically model bonding~\cite{bodei2014linkmembrane}, and can effectively encode compartmentalized calculi such as Mobile Ambients~\cite{bodei2013linkcalculus} and the Brane Calculus~\cite{bodei2014linkmembrane}, but is less symmetric and arguably corresponds less directly with molecular bonding than our mechanism of multiway bonding. The link-calculus also does not have a quantitative semantics and so does not address the issue of assigning rates to multiway interactions. Synchronisation Algebras with Mobility (SAMs)~\cite{lanese2005} and PRISMA~\cite{bruni2007,bruni2008} present a general approach to modelling different synchronisation methods, including multi-way communication and mobility, and in particular, the Hoare SAM~\cite[Example 2]{lanese2005} includes a simultaneous agreement operation based on name fusions~\cite{parrow1998fusion} with some similarities to that of the bond-calculus. Also worth mentioning is GPEPAc~\cite[Chapter~7]{stefanek2014thesis} which extends a variant of PEPA with channels which allows agents to establish sessions and considers fluid analysis by extracting a PCTMC (Population Continuous-Time Markov Chain).

\section{Semantics and simulation}
\label{sec:semantics}
In this section we will attempt to give an indication of the semantics of the language, and how to simulate and analyse models; a full detailed formal treatment of the semantics is contained in the Supplementary Material.

\subsection{Species transition system}

Firstly, we give a formal definition for patterns of sites,
\begin{dfn}
  A \emph{cluster} $\gamma \in \textsc{Cluster}$ consists of a bag $\Lbag s_1, \ldots, s_n \Rbag$ of sites ${s_1,\ldots,s_n \in \textsc{Site}}$, and we write $\gamma = s_1|\ldots|s_n$. A \emph{pattern} $\bm\gamma \in \textsc{Pattern}$ consists of a bag $\Lbag\gamma_1, \ldots, \gamma_n\Rbag$ of clusters ${\gamma_1,\ldots,\gamma_n \in \textsc{Cluster}}$, and we write $\bm\gamma = \gamma_1\apar\ldots\apar\gamma_n$.
\end{dfn}
\begin{figure}
  \begin{align*}
    \begin{matrix}\textsc{Site} \\ s \end{matrix} && \subseteq && \begin{matrix}\textsc{Cluster}\\\gamma = s_1|\ldots|s_n\end{matrix} && \subseteq && \begin{matrix}\textsc{Pattern}\\\bm{\gamma} = \gamma_1 \apar \ldots \apar \gamma_n\end{matrix}
  \end{align*}
  \caption{Sites, clusters, and patterns.}
  \label{patterns}
\end{figure}
That is, patterns are made up of clusters, which are in turn made up of sites (as shown in Figure~\ref{patterns}), and patterns and clusters are represented as bags/multisets.

Next we must consider the behaviour of a single molecule of a given species. To do this we first introduce \emph{abstractions} $F \triangleq (\ell_1,\ldots,\ell_n)A$ which represent a species $A$ with a number of unknown bound locations $\ell_1,\ldots,\ell_n$. Then we consider a prefix $s@m(\ell_1,\ldots,\ell_n).A$ as shorthand for a simpler prefix $s@m.F$ leading to an abstraction $F\triangleq(\ell_1,\ldots,\ell_n)A$. Abstractions represent the products of an open potential reaction and we extend the operations of parallel composition and restriction to abstractions,
\begin{dfn}
  The \emph{parallel composition} or \emph{colocation} of abstractions $(m_1,\ldots,m_p)A$ and $(m_1,\ldots,m_q)B$ is defined by,
  \begin{align*}
  (\ell_1,\ldots,\ell_p)A \para (\ell_1,\ldots,\ell_q)B =
  (\ell_1,\ldots,\ell_s)(A \para B)
  \text{,}
  \end{align*}
  where $s = \max\{p,q\}$. We extend this definition to all abstractions by identifying abstractions upto $\alpha$-renaming.
\end{dfn}
\begin{dfn}
  The \emph{restriction} of names $\ell_1,\ldots,\ell_p$ in an abstraction $(m_1,\ldots,m_q)A$ is defined by,
  \begin{align*}
  (\nu\,\ell_1,\ldots,\ell_p)(m_1,\ldots,m_q)A
  =
  (m_1,\ldots,m_q)(\nu\,\ell_1,\ldots,\ell_p)A
  \end{align*}
  where location names $\ell_1,\ldots,\ell_p$ and $m_1,\ldots,m_q$ are assumed to be distinct by $\alpha$-renaming.
\end{dfn}
Parallel composition allows an abstraction $F$ representing the products of a reaction to be expanded as more reactants join the reaction. Once all the reactants have joined, an abstraction can be \emph{committed}, giving the resulting species.
\begin{dfn}
  The \emph{committed product} of an abstraction $F$ is the species given by,
  \[
    \mathrm{commit}( (\ell_1,\ldots,\ell_n)A) \triangleq (\nu\,\ell_1,\ldots,\ell_n)A
    .
  \]
\end{dfn}

We can then define an operational semantics for species via a labelled multi-transition system system, with a transition $\!\transl{A}{\gamma}{\ell}{\!F}{1.5}\!$ representing the possibility of a transition from a species $A$ to an abstraction $F$ upon reaction at site cluster $\gamma$ at location $\ell$ (we also allow the location to be the \emph{ambient location} $\top$ preventing further reactions at the species level). The transition system is defined via the rules in Supplementary Material Figure~3; most notable is the communication rule,
\begin{align*}
  \prftree[r]
  {\textsc{Com}}
  {\trans{A}{\gamma}{\ell}{F}}
  {\trans{B}{\delta}{\ell}{G}}
  {\ell \neq \top}
  {\trans{A \para B}{\gamma|\delta}{\ell}{F \para G}}
\end{align*}
which states that two transitions at the same (non-ambient) location $\ell$ may be combined into a composite transition, whence the abstractions are combined by colocation and the site clusters are composed.

\subsection{Structural congruence and prime species}

Before we may define the semantics at the mixture level, we must first specify when two species are equivalent (and so represent a single chemical species) and how a species may be decomposed into a composition of prime species (representing distinct molecules). We define which species are equivalent via a structural congruence relation $\equiv$. The structural congruence $\equiv$ for the bond calculus is largely similar to that of the $\pi$-calculus and is defined by the rules in Supplementary Material Section~1.5; henceforth we identify species/abstractions upto structural congruence.

We can then use this to define a prime species, which indicates when a species cannot be broken down into multiple independent subcomponents (and so represents a single molecule),
\begin{dfn}
  A species $S$ is \emph{prime} if, for all species $A, B$,
  \[
  S \equiv A \para B
  \Rightarrow
  A \equiv \unit \text{ or } B \equiv \unit
  \text{.}
  \]
\end{dfn}
This then allows us to decompose any species $S$ as a unique parallel composition of prime species, and we denote by $\operatorname{primes}(S)$, the bag of prime factors of $S$.
\begin{prop}
  For any species $S$, we have a \emph{unique prime decomposition}. That is, there is a unique bag of prime species $\operatorname{primes}(S) = \Lbag P_1,\ldots,P_n \Rbag$ such that,
  \[
  S \equiv \mathop{\para}\limits_{P \in \operatorname{primes}(S)} P \equiv P_1 \para \ldots \para P_n .
  \]
  \begin{proof}
    This follows by using a confluent and terminating system to rewriting rules to reduce species to the unique normal form defined in Supplementary Material Section~1.5 and is similar to \cite[Theorem~11]{kwiatkowski/stark:continuous-pi} (which is expanded in~\cite[Appendex~A]{marekthesis}).
  \end{proof}
\end{prop}

Now we have decomposed a species into primes, we can also decompose any mixture $\Pi$ into a unique parallel composition of prime species,
\[
  \Pi
  \equiv \alpha_1 S_1 \ppar \ldots \ppar \alpha_n S_n
  \equiv
  \mathop{\ppar}_{\text{$S$ prime}} \conc{S}_\Pi S
\]
Then we see that mixtures form an (infinite dimensional) real vector space.
\begin{dfn}
  The mixture space $(\mathbb M, \cdot, +, \unit)$ is the real vector space of mixtures with addition $+$ given by parallel composition $\ppar$ and scalar multiplication $\gamma \cdot \sum \conc{S}_{\Pi} = \sum \gamma \conc{S}_{\Pi}$.
\end{dfn}

If we define the embedding $\emb{S}$ of a single species as below, then we see that the vectors $\emb{P}$ for prime species $P$ form a basis for $\mathbb M$. 
\begin{dfn}
  The species embedding is the map $\emb{\cdot}:\textsc{Spec} \to \mathbb M$ defined by
  \[
    \emb{S} = 1 \cdot S = \sum_{P \in \operatorname{primes} S} 1 \cdot P.
  \]
\end{dfn}

\subsection{Mixture semantics}

We will now define the dynamics of the system, as a vector field $\frac{\mathrm d \Pi}{\mathrm d t}$, assigning an evolution vector to each mixture $\Pi$. In order to develop our semantics we will need a number of auxiliary vector spaces for patterns, clusters, and transitions allowing us to raise the other elements of our language to the same level as the mixture space.
\begin{dfn}
  We define the following (infinite dimensional) real vector spaces:
  \begin{itemize}
    \item The \emph{pattern space} $\mathbb P = \textsc{Pattern} \to_{\mathrm{fin}} \mathbb R$ is the space of finite real vectors indexed by patterns.
    \item The \emph{cluster space} $\mathbb G = \textsc{Cluster} \to_{\mathrm{fin}} \mathbb R \subseteq \mathbb P$ is the space of finite real vectors indexed by clusters of sites.
    \item The \emph{transition space} $\mathbb T = \textsc{Spec}\times\textsc{Abst} \to_{\mathrm{fin}} \mathbb R$ is the space of finite real vectors indexed by transitions $A \to F \triangleq (A,F)$.
  \end{itemize}
  In each case we require that these vectors are \emph{finitely supported real functions} i.e. each $f:\mathbb X \to_{\mathrm{fin}} \mathbb R$, is a function $f:\mathbb X \to \mathbb R$ such that the set $\operatorname{supp} f = \{ x \in \mathbb X \mathrel : f(x) \neq 0 \}$ is finite.
\end{dfn}
These are given the bases of indicator functions $\I{\gamma_1{\apar}\ldots{\apar}\gamma_n}, \I{\gamma}, \T{A}{F}$ respectively, where, the indicator functions $\I{X}:\mathbb X \to \mathbb R$  are defined such that $\I{X}{Y} = 1$ if $X = Y$ and $0$ otherwise. The semantics for mixtures is defined in terms of linear maps $M \in \mathcal L(\mathbb P, \mathbb T)$ which map pattern vectors into transition vectors. We will also use the basis matrices for $\mathcal L(\mathbb P,\mathbb T)$ given as the linear maps $\E{A}{F}{\gamma} \in \mathcal L(\mathbb P, \mathbb T)$ defined such that $\E{A}{F}{\gamma}\I{\gamma} = \I{A \to F}$ if $\gamma = \delta$ and $\mathbf 0$ otherwise. We will call finite sums of these basis maps (finite) matrices $M \in \mathcal M(\mathbb P,\mathbb T)$ (these are infinite dimensional matrices with finitely many nonzero entries, or, more formally, finite rank operators~\cite{abstractanalysis}). Since there are infinitely many potential complexes of species our semantics is specified in terms of infinite dimensional abstract vector spaces~\cite{lang1987linear} (as in \cite{kwiatkowski/stark:continuous-pi}), however, since we only allow mixtures of finitely many species, all of the vectors we consider will have only finitely many non-zero entries and be treated as ordinary column vectors.

We can now define the \emph{transition matrix} of a mixture $\Pi$ as follows.
\begin{dfn}
  The \emph{transition matrix} $\mathcal T(\Pi) \in \mathcal M(\mathbb P,\mathbb T)$ is defined for any mixture $\Pi$ by
  \begin{align*}
    \mathcal T (\alpha A)
      & \triangleq
        \alpha
        \sum\multiset{
          \E{S}{F}{\gamma}
          \mathrel{:}
          S \in \mathrm{primes}(A),
          \!\transl{S}{\gamma}{\top}{\!F}{1.5}\!
        } \\
    \mathcal T(\Pi \ppar \Phi)
      & \triangleq
      \mathcal T(\Pi) + \mathcal T(\Phi)
  \end{align*}
\end{dfn}

We note that for a finite mixture $\Pi \in \mathbb M$, this is indeed a matrix $\mathcal T(\Pi) \in \mathcal M(\mathbb P,\mathbb T)$. This only records basic interactions from a single prime species $S$ which are labelled by site clusters $\bm \gamma$. To derive $n$-way interactions we introduce the $\emph{interaction tensor}$ $\odot$.
\begin{dfn}
  The \emph{interaction tensor} is the bilinear map $\odot:\mathcal M(\mathbb{P}, \mathbb{T}) \times \mathcal M(\mathbb{P}, \mathbb{T}) \to \mathcal M(\mathbb P, \mathbb T)$ defined by,
  \begin{align*}
    \E{A}{F}{\bm\gamma} \odot \E{B}{G}{\bm\delta}
    \triangleq
      \E{A|B}{F|G}{\bm\gamma{\parallel}\bm\delta}
  \end{align*}
\end{dfn}
Then we may find all possible multiway interactions matching a given pattern $\bm \gamma$, by expanding $\mathcal T(\Pi)$ to the linear map $\exp_{\odot}(\mathcal T(\Pi)) \in \mathcal L(\mathbb P,\mathbb T)$ defined by
\begin{align*}
  \exp_{\odot} (\mathcal T(\Pi))
  & \triangleq
  \sum_{n = 0}^\infty \frac{1}{n!} \mathcal T(\Pi)^{\odot n}
  =
     \mathbf 1
   + \mathcal T(\Pi)
   + \frac 1 {2!} (\mathcal T(\Pi) \odot \mathcal T(\Pi))
   + \frac 1 {3!} (\mathcal T(\Pi) \odot \mathcal T(\Pi) \odot \mathcal T(\Pi))
   + \ldots
\end{align*}
where addition of maps is defined pointwise and $\mathcal T(\Pi)^{\odot 0} = \mathbf 1 = \E{\unit}{\unit}{\varnothing}$. This gives a well defined linear map, since for any basis pattern vector $\I{\bm \gamma}$, we get a finite sum
\[
  \exp_{\odot} (\mathcal T(\Pi))\I{\bm \gamma}
  =
  \sum_{n = 0}^\infty \frac{1}{n!} \mathcal T(\Pi)^{\odot n}\I{\bm \gamma}
  =
  \frac{1}{\abs{\bm \gamma}!}
  \mathcal T(\Pi)^{\odot \abs{\bm \gamma}} \I{\bm \gamma}.
\]
where $\abs{\bm \gamma}$ is the size of $\bm \gamma$, since $\mathcal T(\Pi)^{\odot n} \I{\bm \gamma} = \mathbf 0$ whenever $n \neq \abs{\bm \gamma}$ (this equation may be taken as an equivalent definition of the linear map $\exp_{\odot}(\mathcal T(\Pi))$ avoiding infinite sums). The factor $1/\abs{\bm \gamma}!$ corrects for the number of times each transition is overcounted by $\odot$ due the commutativity of $\ppar$.

We still need to calculate the rates of each reaction, which depend on the concentrations of each cluster of sites $\gamma$. First we define the vector of site cluster concentrations in $\Pi$ as,
\begin{dfn}
The \emph{concentration of cluster $\gamma \in \mathbb G$ in mixture $\Pi$} is the real number $\conc{\gamma}_\Pi$ defined by
\begin{align*}
  \conc{\gamma}_{\Pi} = \onenorm{\mathcal T(\Pi) \I{\gamma}}
\end{align*}
where the $\ell^1$-norm $\onenorm{\cdot}$ is defined as $\onenorm{\sum_{j} \alpha_j \T{A_j}{F_j}} = \sum_j \abs{\alpha_j}$. We then define the \emph{site concentration vector of $\Pi$} by,
\[
  \mathcal C(\Pi) \triangleq \sum_{\gamma \in \textsc{Cluster}} \conc{\gamma}_\Pi\I{\gamma}
  \in \mathbb G
  .
\]
\end{dfn}
Next we define the \emph{rate vector} $\mathcal R_{\mathcal A}$ as a function of the global site concentrations. This contains the stochiometric rates for each pattern $\bm \gamma$ in an affinity network $\mathcal A$, and provides a compositional semantics for affinity networks.
\begin{dfn}
  The \emph{rate vector $\mathcal R_{\mathcal A}:\mathbb G \to \mathbb{P}$} for affinity network $\mathcal A$, is the (non-linear) function defined by
  \begin{align*}
    \mathcal R_{\mathcal A}\!\left(\sum_{\gamma \in \textsc{Cluster}}\conc{\gamma}\I{\gamma}\right)
    & = \sum_{(\gamma_1\apar\ldots\apar\gamma_m \atrate f) \in \mathcal A}
    \!\left(\frac{f(\conc{\gamma_1},\ldots,\conc{\gamma_m})}{\conc{\gamma_1}\ldots\conc{\gamma_m}}\right)\!
    \I{\gamma_1{\apar}\ldots{\apar}\gamma_m}.
  \end{align*}
\end{dfn}

\subsection{Dynamics}

The dynamics of the system may now be defined as a vector field over mixtures, defining an instantaneous evolution vector $\frac{\mathrm d \Pi}{\mathrm d t}$ for any mixture $\Pi$ as follows.
\begin{dfn}
  The \emph{evolution vector $\frac{\mathrm d \Pi}{\mathrm d t} \in \mathbb M$ of a mixture $\Pi$ under affinity network $\mathcal A$} is defined by,
  \begin{align*}
    \frac{\mathrm d \Pi}{\mathrm d t}
    & \triangleq
    \mathcal F \exp_{\odot}(\mathcal T(\Pi))\mathcal R_{\mathcal A}(\mathcal C(\Pi)).
  \end{align*}
  where the \emph{finalisation map} $\mathcal F \in \mathcal L(\mathbb T, \mathbb M)$ is the linear map defined by 
  \[
    \mathcal F \T{A}{F} = \emb{\mathrm{commit}(F)} - \emb{A} .
  \]
\end{dfn}

Whilst it possible for this vector field to be infinite dimensional (for instance, in polymerisation reactions~\cite{cardelli2009actin,stefanek2014thesis}), in most cases given a fixed starting mixture $\Pi_0$, all mixtures which may be reached are a linear combination of a finite set of prime species $P_1,\ldots,P_n$ and so we may derive the evolution vector, $\frac{\mathrm d \widehat{\Pi}}{\mathrm d t} = \mathbf f(\conc{P_1},\ldots,\conc{P_n})$ of a general symbolic mixture $\widehat{\Pi} = \conc{P_1} P_1 \ppar \ldots \ppar \conc{P_n} P_n $ and hence extract a system of coupled differential equations capturing the dynamics of the system as described in Supplementary Material Section~3.3. It is possible to perform stochastic simulation of a system by extracting a chemical reaction network with reactants representing prime species and rates derived via the semantics as described in Supplementary Material Section~3.1. Our implementation of the bond-calculus supports both ODE extraction and numerical simulation using SciPy~\cite{scipy} or stochastic simulation via StochPy~\cite{stochpy2013}. 

\section{Case study: Kuznetsov's model of immunogenic tumour growth}
\label{sec:modellingexamples}

\begin{figure}[t]
  \centering
  \begin{subfigure}{0.5\textwidth}
    \centering
    \includegraphics[width=\textwidth]{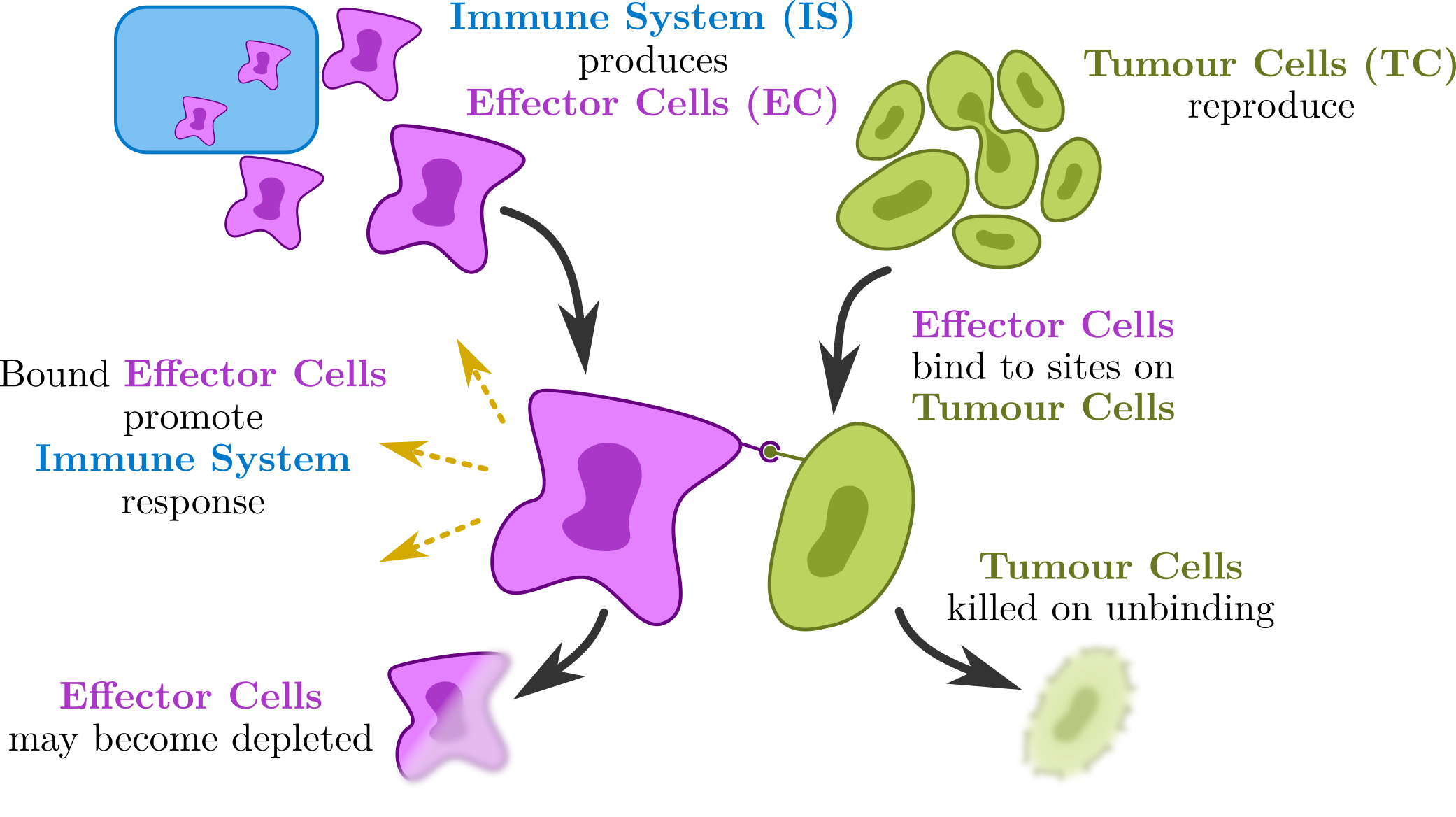}
    \caption{Immune interaction model.}
    \label{fig:immuneinteractions}
  \end{subfigure}%
  \begin{subfigure}{0.5\textwidth}
    \centering
    \schemestart
      EC \+ TC
      \arrow{<=>[$k_1$][$k_{-1}$]}
      EC--TC
      \arrow(c--ec){->[*{0}$k_2$]}[30]
      EC
      \arrow(@c--ec){->[][*{0}$k_3$]}[-30]
      TC
    \schemestop
    \caption{\emph{In vitro} reaction scheme.}
    \label{fig:kuznetsov-reactions}
  \end{subfigure}
  \caption{Kuznetsov's tumour immune model.}
\end{figure}
Mathematical modelling is a significant part of efforts to better understand how tumours develop, to improve treatments, and are increasingly being developed to predict patient specific treatment responses; one survey is~\cite{enderling2014modellingtumourgrowth}. Whilst many tumours evolve mechanisms to evade immune response to various degrees, immune interactions still play a significant part in understanding tumour growth and treatments, with the effectiveness of treatments often depending on the dynamics of these interactions. Furthermore, immunotheropy seeks to devise new treatments which enhance the immune response, and has recently shown promising results in a range of clinical settings~\cite{farkona2016immunotherapy,topalian2011cancer,sharma2015immune}. Therefore there has been a lot of interest in modelling the dynamics of tumour immune interactions. One of the most influential models of this type is Kuznetsov's model of immunogenic tumour growth~\cite{kuznetsov1994tumourmodel} which models the growth of tumour cells and immune effector cells as two species, and the effect of the interactions between them. Kuznetsov's classic model illustrates the key features of tumour immune interactions, has been fitted against experimental data in chimaeral mice~\cite{kuznetsov1994tumourmodel}, and has been extended by many subsequent mathematical models~\cite{kirschner1998,villasana2003,galach2003,depillis2005,donofrio2005}.

At the core of the model is the interaction between Tumour Cells ($\mathrm{TC}$), immune Effector Cells ($\mathrm{EC}$) (e.g. Cytotoxic T Lymphocytes or Natural Killer cells), and the wider Immune System (IS) as shown in Figure~\ref{fig:immuneinteractions}. The Effector Cells bind to the Tumour Cells triggering one of three possible outcomes: the TC and EC unbind leaving both undamaged, the EC unbinds destroying the TC, or the TC unbinds destroying/depleting the EC. Additionally, the bound $\ch{EC-TC}$ pairs can signal the Immune System (IS) to send more Effector Cells. This model is captured in bond-calculus via the following agents, which represent the bound and unbound states of the $\mathrm{TC}$/$\mathrm{EC}$ along with an agent representing the Immune System IS,
\begin{align*}
  \mathrm{TC} & \triangleq
    \mathrm{growTC}.(\mathrm{TC} \para \mathrm{TC}) +
    \mathrm{bindTC}(\ell).\mathrm{TC}^*_\ell \\
    & + \mathrm{limitImmune}.\mathrm{TC}
    + \mathrm{consumeResources}.\mathrm{TC}
    \\
  \mathrm{TC}^*_\ell & \triangleq
    \mathrm{unbindTC}@\ell.\mathrm{TC} +
    \mathrm{dieTC}@\ell.\unit \\
    & + \mathrm{consumeResources}.\mathrm{TC^*_\ell} \\
  \mathrm{EC} & \triangleq
    \mathrm{bindEC}(\ell).\mathrm{EC}^*_\ell
    + \mathrm{dieEC}.\unit
    \\
  \mathrm{EC}^*_\ell & \triangleq
    \mathrm{unbindEC}@\ell.\mathrm{EC} +
    \mathrm{dieEC}@\ell.\unit \\
    & + \mathrm{pathogenDetected}.\mathrm{EC}^*_\ell \\
  \mathrm{IS} & \triangleq
    \mathrm{spawnEC}.(\mathrm{IS} \para \mathrm{EC}).
\end{align*}
The interactions between the various types of agents are described by the affinity network,
\begin{align*}
  \mathcal A \triangleq \big\{\,
    & \mathrm{spawnEC} \apar \mathrm{pathogenDetected} \apar \mathrm{limitImmune}
      \atrate \operatorname{Response}_{s,f,g},\, \\
    & \mathrm{bindTC} \apar \mathrm{bindEC} \atrate \operatorname{MA}_{k_1},\, 
      \mathrm{growTC} \apar \mathrm{consumeResources}
      \atrate \operatorname{Logistic}_{a,b},\, \\
    & \mathrm{unbindTC}|\mathrm{unbindEC} \atrate \operatorname{MA}_{k_{-1}},\,
      \mathrm{unbindTC}|\mathrm{dieEC} \atrate \mathrm{MA}_{k_2},\,\\
    & \mathrm{dieTC}|\mathrm{unbindEC} \atrate \operatorname{MA}_{k_{-1}},\,
      \mathrm{dieEC} \atrate \mathrm{MA}_{k_2}
  \,\big\}.
\end{align*}
For example, the affinity pattern $\mathrm{bindTC} \apar \mathrm{bindEC} \atrate \operatorname{MA}_{k_1}$ specifies that unbound TCs and ECs may interact on the $\mathrm{bindTC}$ and $\mathrm{bindEC}$ sites respectively to become bound, forming the dynamic complex,
\[
  \ch{EC-TC} \triangleq (\nu\,\ell)\Bigl(\ch{EC}^*_{\ell} \para \ch{TC}^*_{\ell}\Bigr)
\]
whilst the affinity patterns specified that $\mathrm{spawnEC} \apar \mathrm{pathogenDetected} \apar \mathrm{limitImmune} \atrate \operatorname{Response}_{s,f,g}$ that the Immune System communicates with both Effector Cells (on the $\mathrm{pathogenDetected}$ site) and Tumour Cells (on the $\mathrm{limitImmune}$ site) to determine the overall rate of effector cell production, according to the general kinetic law,
\begin{align*}
  \operatorname{Response}_{s,f,g}(\conct{spawnEC},\conct{pathogenDetected},\conct{limitImmune})
    & \triangleq
    s + \frac{f\conct{pathogenDetected}}{g + \conct{limitImmune}}
\end{align*}
  Finally, the rate of Tumour Cell growth is determined by the affinity pattern $\mathrm{growTC} \apar \mathrm{consumeResources} \atrate \operatorname{Logistic}_{a,b}$ with the logistic law defined by
\begin{align*}
  \operatorname{Logistic}_{a,b}(\conct{growTC},\conct{consumeResources})
    & \triangleq
    a \conct{growTC}(1 - b \conct{consumeResources})
  .
\end{align*}
This means that whilst the growth rate depends on the current concentration of (unbound) tumour cells, it is limited by the available resources up to a carrying capacity $1/b$ -- since the function rate is determined by site concentrations rather than species concentrations, the growth rate is determined by the concentration of only the unbound TC cells (which possess the $\mathrm{growTC}$ site), whilst both bound and unbound TC cells contribute to the resource consumption (since both possess the $\mathrm{consumeResources}$ site).
\begin{figure}
  \centering
  \includegraphics[width=\textwidth]{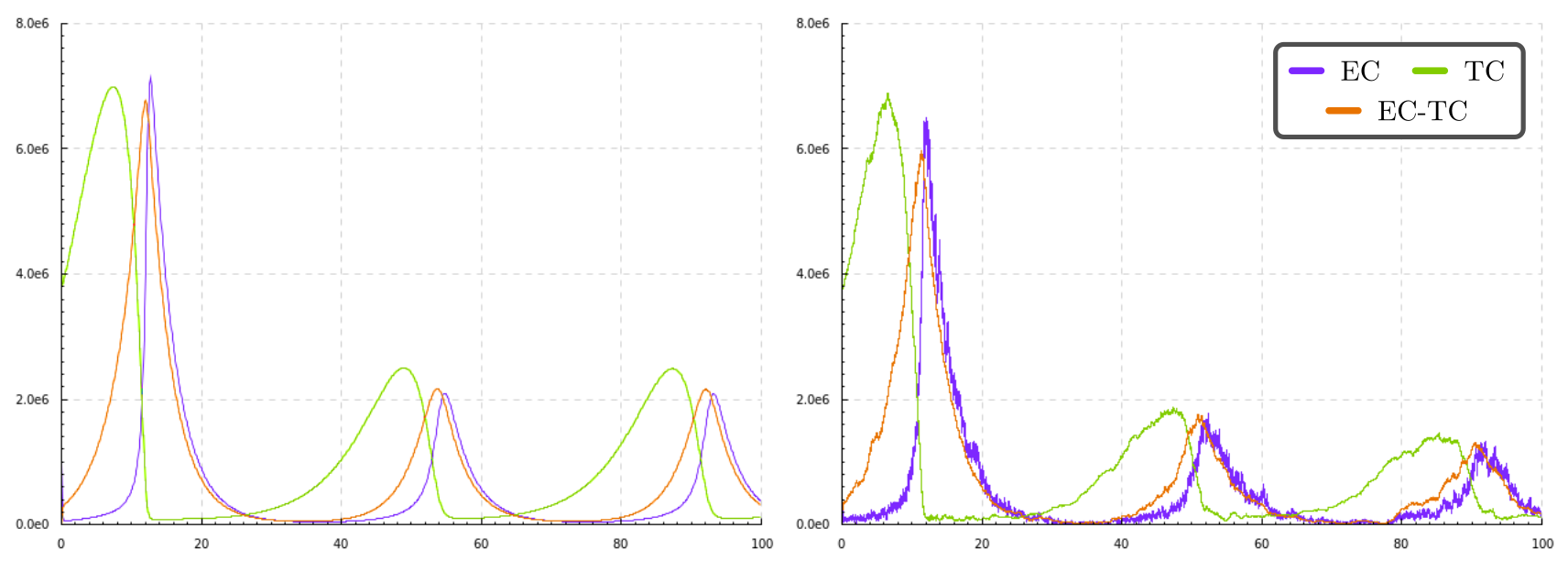}
  \caption{A sample run of Kuznetsov's model, showing the possibility of sustained oscillations, with simulated ODEs on the left, and stochastic simulation with discretization step-size $h = 10000$ on the right.}
  \label{fig:kuznetsov_mass_action}
\end{figure}

From this model we can derive the following system of differential equations,
\begin{align*}
  \frac{\mathrm d \conch{EC}}{\mathrm d t} & =
    s + \frac{f [\ch{EC-TC}]}{g + \conch{TC}} - d_1 \conch{EC} - k_1 \conch{EC}\!\conch{TC} + (k_{-1} + k_2) [\ch{EC-TC}] \\
  \frac{\mathrm d \conch{TC}}{\mathrm d t} & =
    a \conch{TC}(1 - b(\conch{TC} + [\ch{EC-TC}])) - k_1 \conch{EC}\!\conch{TC} + (k_{-1} + k_3) [\ch{EC-TC}] \\
  \frac{\mathrm d [\ch{EC-TC}]}{\mathrm d t} & =
    k_1 \conch{EC}\!\conch{TC} - (k_{-1} + k_2 + k_3) [\ch{EC-TC}]
\end{align*}
which matches the mass action version of Kuznetsov's model~\cite{kuznetsov1994tumourmodel}. The results of simulating the model for one particular set of parameter values is shown in Figure~\ref{fig:kuznetsov_mass_action}~(left) -- this shows oscillatory behaviour, demonstrating the possibility for tumour regrowth in the case the immune response is too slow or insufficiently sustained. It is also possible to perform stochastic simulation by splitting the continuous concentrations as discrete levels with a given discretization step size $h$ and adjusting the rates accordingly (similarly to the procedure described for Bio-PEPA in~\cite{biopepapaper}). A sample stochastic simulation run is shown in Figure~\ref{fig:kuznetsov_mass_action} (performed via StochPy~\cite{stochpy2013} using Gillespie's Stochastic Simulation Algorithm), however, stochastic simulation with $h = 1$ (i.e. with separate levels for each cell of the population) is not feasible given the population sizes involved in the model. Whilst under many cases the behaviour of the ODEs will approximate the mean stochastic behaviour of the system given sufficiently large agent populations~\cite{geisweiller2008continousdiscretemodels,tribastone2012scalable}, the dynamics of the stochastic model may include features impossible in the ODEs (for example, the eventual extinction of the tumour cells once the population reaches $0$), and the validity of both the approximation (and indeed of stochastic simulation) depends on the kinetic law used in the model (see \cite[Section~8.3]{biopepapaper}). This highlights that, as always, care should be taken when using continuous methods such as differential equations to model fundamentally stochastic phenomena such as cellular interactions, especially when there is a possibility of small populations.

\section{Discussion}

Since the first biological models using the $\pi$-calculus~\cite{regev2001,ciobanu2000formal}, a succession of process algebras have been applied to better describe different aspects of biochemical systems, including quantitative rates~\cite{calder2006pepapathway,calder2006stronger,phillips2007stochastic}, general kinetics laws~\cite{biopepapaper,Bortolussi2008sccpbio,Dematte2008BlenXTutorial}, and multi-way interactions~\cite{ciocchetta2007biologicaltransactions,ciocchetta2008blenXT,biopepapaper,Bortolussi2008sccpbio,bodei2014linkmembrane} and to find more expressive communication operations to simplify the modelling process~\cite{Priami2005betabinders,Dematte2008BlenXTutorial,kuttler2007spico,versari2008piat,John2008attributedpi,bodei2014linkmembrane,ciobanu2008}. The variety of process calculi is almost matched by the variety of techniques which have been applied to simulating and analysing models~\cite{biopepapaper,caravagna2012biopepad,galpin2013hype,slegers2010langevin,clark2010verifyingbiopepa}. In particular, the continuous interpretation of models as employed by continuous $\pi$~\cite{kwiatkowski/stark:continuous-pi}, Bio-PEPA~\cite{biopepapaper,tribastone2012scalable}, and $\mathcal L\pi$~\cite{stefanek2009,stefanek2009spatial}, makes it possible to extract systems of ODEs from process algebra models, allowing them to be efficiently simulated and validated against existing mathematical models. This body of work has made significant progress towards showing biological systems can be effectively analysed as communicating systems, and that formal models are an effective tool to study their behaviour. However, the intricate and varied forms of interaction in biological systems remains a persistent challenge to creating models which adequately describe the system whilst remaining comprehensible to the human reader, and the complexity of existing frameworks poses a significant barrier to the wider adoption of formal modelling in biology.

In this paper we presented the bond-calculus, a new process algebra which attempts to capture the full behaviour of biochemical reactions including general kinetic laws, dynamic bonding, and multi-way interactions. To do this we have integrated ideas from many existing process calculi~\cite{kwiatkowski/stark:continuous-pi,biopepapaper,Dematte2008BlenXTutorial,bodei2013linkcalculus,versari2008piat,sangiorgi1996pi,fournet2002joincalculus,kuttler2007spico,danos2004formal}, whilst introducing some ideas of our own including affinity patterns which provide an easy way to specify compatibility of sites through pattern matching, and symmetric multi-way bonding, a more natural communication operation for $n$-way bonding. A major challenge of this work was integrating all of these features into a single harmonious whole, however, the resulting language manages to retain much of the parsimonious appeal of the $\pi$-calculus, whilst different features such as multi-way bonding, locations, affinity patterns, and general kinetic laws complement each other to effectively describe complex biochemical reactions. All of this rests upon the continuous semantics in Section~\ref{sec:semantics} which defines a compositional representation of mixtures and affinity networks, whilst supporting extraction of ODEs and chemical reaction networks. This significantly extends the semantics of continuous $\pi$~\cite{kwiatkowski/stark:continuous-pi} which was restricted to binary and unary mass action reactions, whilst retaining the key property of compositionality. 

We hope that this language can be applied to improve and simplify existing biological models, whilst opening the door to new models exploiting the key features of the language such as multi-way bonding. Additional work will be required to extend the theoretical frameworks of existing binary~\cite{milner1999communicatingandmobilesystems} and multi-way~\cite{Hoare2002,fournet2002joincalculus,biopepapaper,bodei2013linkcalculus} process calculi to the bond-calculus and to clarify its relation to these calculi. Whilst the bond-calculus supports ODE extraction and stochastic simulation (via StochPy~\cite{stochpy2013}), there are a wide range of other simulation and analysis techniques such as hybrid simulation and model checking which could be applied to provide more insight into models. In the future we also hope to apply our semantics to develop new methods of analysing systems which make full use of the detailed compositional view it gives us of their behaviour.

\bibliography{main}{}
\bibliographystyle{plain}

\end{document}